\documentclass[10pt,conference]{IEEEtran}
\usepackage{amsmath,amssymb,amsfonts}
\usepackage{algorithmic}
\usepackage{graphicx}
\usepackage{textcomp}
\usepackage{xcolor}
\usepackage[numbers,sort&compress]{natbib}
\def\BibTeX{{\rm B\kern-.05em{\sc i\kern-.025em b}\kern-.08em
T\kern-.1667em\lower.7ex\hbox{E}\kern-.125emX}}
\usepackage[hidelinks]{hyperref}

\usepackage[numbers]{natbib}
\usepackage[inline]{enumitem}
\usepackage{url}

\definecolor{darkgreen}{rgb}{0,0.6,0}

\begin{document}

\title{Addressing Research Software Sustainability via Institutes}

\author{
\IEEEauthorblockN{Daniel S. Katz\IEEEauthorrefmark{1}, Jeffrey C. Carver\IEEEauthorrefmark{2}, Neil P. Chue Hong\IEEEauthorrefmark{3}, Sandra Gesing\IEEEauthorrefmark{4},\\ Simon Hettrick\IEEEauthorrefmark{5}, Tom Honeyman\IEEEauthorrefmark{6}, Karthik Ram\IEEEauthorrefmark{7}, and Nicholas Weber\IEEEauthorrefmark{8}} 
\IEEEauthorblockA{\IEEEauthorrefmark{1}NCSA \& CS \& ECE \& iSchool, University of Illinois, Urbana, IL, USA, ORCID: 0000-0001-5934-7525} 
\IEEEauthorblockA{\IEEEauthorrefmark{2}Department of Computer Science, University of Alabama, Tuscaloosa, AL, USA, Email: carver@cs.ua.edu}
\IEEEauthorblockA{\IEEEauthorrefmark{3}Software Sustainability Institute \& EPCC, University of Edinburgh, Edinburgh, UK, ORCID: 0000-0002-8876-7606}
\IEEEauthorblockA{\IEEEauthorrefmark{4}Center for Research Computing, University of Notre Dame, Notre Dame, IN, USA, ORCID: 0000-0002-6051-0673}
\IEEEauthorblockA{\IEEEauthorrefmark{5}Software Sustainability Institute \& ECS, University of Southampton, Southampton, UK, ORCID: 0000-0002-6809-5195}
\IEEEauthorblockA{\IEEEauthorrefmark{6}Software program, Australian Research Data Commons, Univ. of Technology, Sydney, Aus., ORCID: 0000-0001-9448-4023}
\IEEEauthorblockA{\IEEEauthorrefmark{7}Berkeley Institute for Data Science, University of California, Berkeley, CA, USA, ORCID: 0000-0002-0233-1757}
\IEEEauthorblockA{\IEEEauthorrefmark{8}Information School, University of Washington, Seattle, WA, USA, ORCID: 0000-0002-6008-3763}
}

\maketitle

\begin{abstract}
Research software is essential to modern research, but it requires ongoing human effort to sustain: to continually adapt to changes in dependencies, to fix bugs, and to add new features. Software sustainability institutes, amongst others, develop, maintain, and disseminate best practices for research software sustainability, and build community around them. These practices can both reduce the amount of effort that is needed and create an environment where the effort is appreciated and rewarded. The UK SSI is such an institute, and the US URSSI and the Australian AuSSI are planning to become institutes, and this extended abstract discusses them and the strengths and weaknesses of this approach.
\end{abstract}

\begin{IEEEkeywords}
software sustainability institutes, software best practices, community, research software, software sustainability
\end{IEEEkeywords}


\section{Introduction and context}

A large fraction of modern research depends on research software~\cite{UK-software-survey, software-in-nature}. 
This software often can be considered to fit into a stack, as described by Hinsen~\cite{software-collapse}, consisting of, from top to bottom: project-specific code (e.g.,  scripts, workflows), domain-specific tools (e.g., community codes), scientific software infrastructure (e.g., math and I/O libraries), and non-scientific software infrastructure (e.g., compilers, standard libraries), all on top of the operating system (OS) and hardware.

The fact that this software works at a given time does not mean that it will work in the future, because the lower layers upon which the software depends (including the OS and hardware) will change over time, and this can cause the software to break. Additionally, there may be bugs discovered in the software that should be fixed, and in many cases, new use cases will arise that require modifying the software. In all cases, there is work needed to keep the possibly fixed or expanded software running correctly on the possibly updated lower layers. We use sustainable to refer to situations where this work is planned for and likely to occur, though in some sense, sustainability is something that can only be evaluated in hindsight; sustainability in the future is merely a prediction.

While some of this software, particularly in the lower layers, is reasonably well-supported by funding agencies, industry, or open source communities, and is likely to be sustained, much of it is also developed by itinerant laborers (graduate students and postdocs) as a side effort in their research activities, and even when a person dedicates long term effort to a software package, at some point they will retire or move on.

\section{Challenges}

Better sustaining of research software can be addressed in multiple ways, including by reducing the work needed to sustain it, and increasing the resources available to do so. To reduce the work, research is needed to develop, maintain, and disseminate best practices and effective tools. To increase the resources, which are often people, since developing and maintaining software is human-intensive, we must increase human effort, either by paying people to work on software or finding other ways to encourage them to choose to do it.

While there is a technical aspect to these two factors (e.g., determining best practices, developing tools), many issues are partially or entirely social (e.g., disseminating best practices and tools, incentivizing contributions to software via hiring and promotion, funding policies that plan for ongoing maintenance), thus community activity is needed to address them.

\section{Existing solutions}


These challenges can be addressed at a global or national scale, and because research funding is typically national, national efforts are arising. Three national activities have been working on this problem in their own countries, as part of a larger community of researchers interested in this problem.

In the UK, the Software Sustainability Institute (SSI, \href{https://www.software.ac.uk/}{software.ac.uk}) was established in 2010 with EPSRC funding to identify key issues and best practices for research software~\cite{SSI_2013}. It has worked with 70+ research groups to improve their software practices and published over 80 guides (some used by $>$50,000 people) ranging from software engineering to managing projects to building communities. Other successes include the SSI's Fellowship program, a cohort of 150 advocates for improving software practice (including diversity and recognition) in their research domains~\cite{SSI_Fellows_2018}, and a partnership with The Carpentries  that developed a 350+ instructor base and delivered training for 7,500 learners at 50+ organizations. Consequences include: an increased use of software engineering practices that improve sustainability, such as version control; software outputs from research projects are more likely to be developed as open source and deposited in a digital repository.

The SSI was instrumental in establishing the Research Software Engineering movement, from the term Research Software Engineer being coined at a 2012 SSI workshop~\cite{RSE_2012} to providing \emph{backbone organization} support to establish the UK RSE Association in 2014 and Society of Research Software Engineering in 2019. This included working with policymakers to develop funding programs such as 
RSE Fellowships~\cite{RSE_Fellowships}
and guidance on recognizing software as a research output in 
the UK Research Evaluation Framework.
There are now RSE associations on three continents, 370+ professional members, and thousands engaged in the wider RSE community. UK researchers are more likely to explicitly budget software engineering effort on research grants.

In the US, a planning project started in 2017 to design a US Research Software Sustainability Institute (URSSI, \href{https://urssi.us}{urssi.us}) and to build community support for it~\cite{URSSI}. The project has run multiple workshops and a large survey to understand US researchers' needs and to determine possible activities to support them. In addition, it has used a set of community activities (e.g., website, blog posts) to build community awareness. These activities and examining the UK SSI's results led to a plan for a US institute: \href{https://plan.urssi.us}{plan.urssi.us}. This plan includes work in community and outreach, education and training, incubation, and policy. The URSSI team is now seeking funding to turn its vision, or at least parts of it, into reality.

\begin{figure}[ht]
    \centering
    \includegraphics[width=0.45\textwidth]{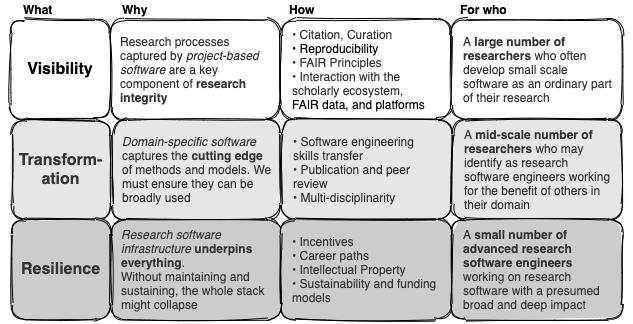}
    \caption{Framework for AuSSI agenda. (FAIR = findable, accessible, interoperable, reusable) \label{fig:AuSSI}}
\end{figure}

In Australia, work began in 2020 to draft a national agenda for research software (agenda framework shown in Figure~\ref{fig:AuSSI}), with the aim of achieving recognition of software as a first class research output. This agenda is heavily informed by the work of the UK SSI and URSSI. Establishing a software sustainability institute, AuSSI, will be validated with the community in 2021. AuSSI will serve to coordinate activities arising from the agenda, and will be housed within, and complement existing activities at, the Australian Research Data Commons (ARDC). The agenda suggests activities in infrastructure, guidance and outreach, community building and advocacy.

\section{Strengths and weaknesses}


As a solution to making research software more sustainable, institutes have some strengths:
\begin{itemize}
    \item Focus on academia and national labs (a government funded activity can be most successful aimed at government-funded institutions)
    \item Collaboration between institutes and communities (institutes and individuals can work together under a common vision to find common or customizable solutions)
    \item Leveraging research-specific concerns such as reproducibility and open science / open research 
\end{itemize}

\noindent and some weaknesses: 
\begin{itemize}
    \item Hard to impact industry (industry is much larger and has different social practices and incentives)
    \item Limited funding and limited time (government funded activities are dependent on government funding, which is generally of fixed duration)
\end{itemize}

Institutes are not a complete solution, but in cooperation with other activities (and with each other), they can help communities change practices, leading to more sustainable research software.

\bibliographystyle{IEEEtranDOI}
\bibliography{bokss-2021-software-institutes.bib}

\end{document}